\documentclass[a4paper,10pt,reqno]{article}
\usepackage{amsmath,amssymb,amsfonts} 
\usepackage{graphics} 
\usepackage{color} 
\usepackage{graphicx,subfigure}
\usepackage[english]{babel} 
\usepackage{empheq}
\usepackage{fancyhdr}
 \usepackage{authblk}
\usepackage{amsmath}

\newcommand{\ds} {\displaystyle}
\providecommand{\abs}[1]{\lvert#1\rvert}

\newtheorem{obs}{Observation}

\title{Determination and biological application of a time dependent thermal parameter and sensitivity analysis for a conduction problem with superficial evaporation}
\date{}

\author[a,b,c]{Natalia N. Salva  \footnote{Corresponding author: Tel.:+54 294 4445900 (5887).\\ \hspace*{4mm} E-mail address: natalia.salva@yahoo.com.ar \\ \hspace*{4mm} Postal address: Laura 7749, (8400) San Carlos de Bariloche, R\'io Negro, Argentina.}}
\author[d]{Mar\'ia S. Herrera}
\author[e,f]{Andrea Monti Hughes}
\author[a,b,c]{Claudio Padra}
\author[e]{Gustavo A. Santa Cruz}

\affil[a]{CONICET, San Carlos de Bariloche, 8400, Argentina }
\affil[b]{Depto. de Mec\'anica Computacional, GIA, Comisi\'on Nacional de Energ\'ia At\'omica, Av. Bustillo 9500, San Carlos de Bariloche, 8400, Argentina}
\affil[c]{Centro Regional Bariloche, Universidad Nacional del Comahue, Quintal 200, San Carlos de Bariloche, 8400, Argentina}
\affil[d]{CETMIC, CIC-CONICET-CCT La Plata, Camino Centenario y 506, M. B. Gonnet, B1897ZCA, Argentina}
\affil[e]{Dpto. de Radiobiolog\'ia, GAATEN, Comisi\'on Nacional de Energ\'ia At\'omica, Av. Gral. Paz 1499, Buenos Aires, B1650KNA, Argentina}
\affil[f]{CONICET, Godoy Cruz 2290, Buenos Aires, C1425FQB, Argentina}

\begin{document}

\maketitle

\begin{abstract}
A boundary value problem, which could represent a transcendent temperature conduction problem with evaporation in a part of the boundary, was studied to determine unknown thermophysical parameters, which can be constants or time dependent functions. The goal of this paper was elucidate which parameters may be determined using only the measured superficial temperature in part of the boundary of the domain. We formulated a nonlinear inverse problem to determine the unknown parameters and a sensitivity analysis was also performed. In particular, we introduced a new way of computing a sensitivity analysis of a parameter which is variable in time. We applied the proposed method to model tissue temperature changes under transient conditions in a biological problem: the hamster cheek pouch. In this case, the time dependent unknown parameter can be associated to the loss of heat due to water evaporation at the superficial layer of the pouch. Finally, we performed the sensitivity analysis to determine the most 
sensible parameters to variations of the superficial experimental data in the hamster cheek pouch.


\end{abstract}



\section{Introduction}

Diffusion problems have been studied in many different areas, such as heat conduction  \cite{Bankoff}, sediment transport in a river  \cite{Sediment}, the study of the weather \cite{Weather}, or in drying food technology  \cite{bananas,cereales,papas}. The heat transport in a certain object may be modeled through a convection-diffusion differential equation. The boundary conditions usually considered are: a constant temperature (Dirichlet condition), a fixed heat flux (Neumann condition) or a mixed condition, where the heat flux depends on the temperature at the boundary.

In \cite{PM} some thermal inverse problems were considered in a mathematical tumor model. They estimated simultaneously unknown thermophysical and geometrical parameters, through an evolutionary algorithm. All the parameters were constants and no evaporation was considered. 
In \cite{BBB2017} a space-dependent convection parameter was determined through a nonlinear least squares technique, using several temperature measurements at different times in different locations of the domain.
In \cite{SPIE2017}, we observed that exophytic tumors and irradiated tissues, principally those with ulcers, exhibit mass moisture transfer in the tissue-air interphase.

In this work,  we resolved a partial differential problem considering a time-dependent extra term in the mixed condition, which can be associated with a loss of heat due to superficial evaporation. Having measurements of temperature in a section of a particular domain during time, a nonlinear inverse problem was formulated to determine some thermophysical parameters such as the superficial heat loss or the heat conductivity. A sensitivity analysis was also  performed. Finally the parameter determination and the sensitivity analysis were validated using our previous study on tissue temperature responses in the hamster cheek pouch \cite{SPIE2017}. 
\section{Description of the conduction problem}
Let us consider a one dimensional spatial domain $\Omega=[0,X_{\max}]$, and the following boundary value problem:

\begin{subequations} \label{eq:ODEProb}
\begin{align}  
& c_1 \dfrac{\partial T }{ \partial t }(x,t)=c_2  \dfrac{\partial^2 T }{ \partial x^2 }(x,t) +c_3 T(x,t)+c_4, & \forall \;  (x,t) \in \Omega \times [0,t_{\max}] \label{eqn:1}\\[2ex] 
& c_5 \dfrac{\partial T}{\partial x}(x,t)= c_6 T(x,t)+c_7+f(t), &  \forall \;  (x,t) \in \Gamma_u \times [0,t_{\max}] \label{eqn:2}\\[2ex] 
 & T(x,t)=g(t), &  \forall \; (x,t) \in \Gamma_b \times [0,t_{\max}] \label{eqn:3}\\[2ex] 
 & T(x,0) =  h(x) & \forall \;  x \in \Omega    \label{eqn:4}
\end{align}
\end{subequations}

\noindent where  $c_i$ are real constants, $f,g:[0,t_{\max}] \to \mathbb{R}$ and $h:\Omega \to \mathbb{R}$ are real functions, and $\Gamma_u=\{X_{\max}\}, \; \Gamma_b=\{0\}$\footnote{We chose one spatial dimension for simplicity. This procedure can be easily extended to more dimensions without loss of generality.}.

In general, the parameters $c_i, \; i=1..7$, $f,g$ and $h$ are fundamental in using this type of differential equation problem with boundary conditions given by Eqs.  (\ref{eqn:2}), (\ref{eqn:3}) and (\ref{eqn:4}). Note that Eq. (\ref{eqn:2}) includes a time-dependent parameter, $f(t)$, which is defined only in the boundary $\Gamma_u$. In particular, $c_i$  are often approximated, and sometimes their values are just guessed. If there is an experimental measure of the temperature $T(x,t)$  on the boundary $\Gamma_u$, an inverse problem can be formulated defining the following cost function:
\begin{equation}
J(T)=\int_0^{t_{\max}}\int_{\Gamma_u} \abs{T(x,t)-T^* (x,t)}^2 dx dt \,,
\label{eq:E}
\end{equation}
where $T^*(x,t)$ is the measured superficial temperature. Therefore, some parameters can be determined minimizing the cost function $J$. 

The goal of this paper is to find which parameters $c_i$  may be determined using only the measured superficial temperature. First we proposed a method to identify the function $f=f(t)$ such that the solution $T$ of the problem (\ref{eq:ODEProb}) minimizes the cost function (Section \ref{secc:detf}). Then, we introduced a new way of computing a sensitivity analysis of a variable parameter (Section \ref{secc:sa}). As an example, in Section \ref{secc:ba}, we applied this problem to model tissue temperature changes, under transient conditions, in the hamster cheek pouch. In this biological problem, the function $f(t)$ can be associated to the loss of heat due to water evaporation at the superficial layer of the cheek pouch.  Finally, we performed the sensitivity analysis to determine the most sensible parameters to variations of the superficial experimental data (Section \ref{secc:sensApl}).   
\section{Determination of the function $\boldsymbol{f}$} \label{secc:detf}

The goal of this section is to determine the function $f:[0,t_{\max}] \to \mathbb{R}$, assuming that the rest of the parameters are known constant.  
To solve the direct problem (\ref{eq:ODEProb}) we used the Finite Element Method (FEM), meshed the spatial-time\footnote{The mesh used is a triangular uniform mesh, where the nodes are obtained by the Cartesian product of a discretization in time and a discretization in space. Using polynomials of degree one and this particular mesh makes the FEM equivalent to a finite difference scheme.} domain $\Omega \times [0,t_{\max}]$ and obtained the triangulation $ \mathfrak{T}=\{\mathcal{T}_i,i=1...M\}$. 
We define the following finite element spaces:

\begin{equation}
S_\mathfrak{T}=\{ v \in H^1(\Omega \times [0,t_{\max}])/ v \in \mathcal{P}_1(\mathcal{T}), \forall  \; \mathcal{T} \in \mathfrak{T} \}, \mbox{ where } \mathcal{P}_1 \mbox{ is the space of polynomials of degree 1,}
\end{equation}

\begin{equation}
V=\{ v \in S_\mathfrak{T}/ v=g,   \mbox{in } \Gamma_b \times [0,t_{\max}] \; \wedge \; v=h \mbox{ in } \Omega \times \{0\} \} ,
\end{equation}
\begin{equation}
V_0=\{ v \in S_\mathfrak{T}/ v=0,   \mbox{in } \Gamma_b \times [0,t_{\max}] \; \wedge \; v=0 \mbox{ in } \Omega \times \{0\} \} .
\end{equation}

The weak formulation of  problem (\ref{eq:ODEProb}) is the following:

\begin{equation*} \label{eq:PV2}
 (VP): \mbox{Find } T \in V/ \;  a(T,\eta)=l(\eta), \; \forall \eta \in V_0
\end{equation*}

where,

\begin{gather} 
\ds a(u,v)=\underset{\Omega \times [0,t_{\max}]}{\int} c_1 \,  \frac{ \partial u}{\partial t}  v+ c_2 \; \frac{ \partial u}{\partial x}  \frac{ \partial v}{\partial x} - c_3 \, u v \; dx dt   -  \underset{\Gamma_u \times [0,t_{\max}]}{\int} \frac{c_2 \, c_6}{c_5} u v \; dx dt \label{eq:a2}\\ 
\ds \label{eq:l2} l(v)=\underset{\Omega \times [0,t_{\max}]}{\int} c_4 \, v \; dx dt\; + \underset{\Gamma_u \times [0,t_{\max}]}{\int} \frac{c_2}{c_5}  (c_7+f(t)) v\; dx dt \,.
\end{gather}

Let $\{(x_i,t_i), i=1,..,N\}$ be the nodes of the triangulation $\mathfrak{T}$, and  $I \subset \{1,...,M\}$ be the subset of index whose nodes are in the boundary  $\Gamma_u \times  [0,t_{\max}]$. We define the function $f$ as follows,

$$f(t)=\ds \sum_{i \in I} \beta_i \eta_i(X_{\max},t),$$

where $\{\eta_i \}_{i=1..N}$ is the base of the space $V$. 

The inverse problem of determining $f$ consists in obtaining the values of $\{ \beta_i \}_{i \in I}$ such that the solution $T$ of the weak formulation problem is close to the measured temperatures. Let 
$T$ be a solution of $(VP)$. The cost function can be approximated by:

$$J(T)=\underset{\Gamma_u \times [0,t_{\max}]}{\int} (T-T^*)^2 \; dx dt \approx \sum_{j=1}^{N_t} (T(X_{\max},t_j)-T^*(t_j))^2\,.$$ 

We defined the following optimization problem: 

\begin{itemize}
 \item[] {\it Find $B=\{ \beta_i \}_{i \in I}$ such that minimizes the cost function $\displaystyle j(\{ \beta_i \}_{i \in I})=J(T_B)$, where $T_B$ is a solution of $(VP)$ using the values of $B=\{ \beta_i\}_{i \in I}$ to define $f(t)$. }
\end{itemize}

We used the Lagrange Method to determine the variation of $j$ by each $\beta_i$, and obtaining a direction of descent. Suppose that for $r \in I-\{i\}$ the values of $\beta_r$ are fixed, and we vary only the parameter $\beta_i$. We define the following function:
\begin{equation}
\mathcal{L}(u,w,\beta_i)=J(u)+a(u,w)-l_{\beta_i}(w) 
\end{equation}
where $a(\cdot,\cdot)$ y $l(\cdot)$ is defined in (\ref{eq:a2}) and (\ref{eq:l2}), and  the subscript ${\beta_i}$ means that we are using in $f(t)$ the fixed values of $\{ \beta_r\}_{r \in I-\{i\}}$ and the value of $\beta_i$ (which may vary).

Note that if $u_{\beta_i}$ is a solution of (VP) then  $\forall w \in V_0: \, \mathcal{L}(u_{\beta_i},w,\beta_i)=J(u_{\beta_i})=j(\beta_i)$, and therefore their derivatives are equal.

Using the chain rule we obtained the derivative of the Lagrangian $\mathcal{L}$ respect to each $\beta_i$:
\begin{equation} \label{eq:derivLs}
 \frac{\delta \mathcal{L}(u_{\beta_i},w_{\beta_i},\beta_i)}{\delta \beta_i}= \underset{\Gamma_u \times [0,t_{\max}]}{\int}  \eta_i w_{\beta_i}  \; dx dt
\end{equation}
where $w_{\beta_i} \in V(\Omega)$ is the adjoint state, which is defined as the unique solution to  the following variational problem:
$$(VPA) \left\{ \begin{array}{ll}
\displaystyle a(v,w_{\beta_i})=-2 \underset{\Gamma_u \times [0,t_{\max}]}{\int} (u_{\beta_i}-T^*) v \; dx dt,& \quad \forall v \in V\\
 w_{\beta_i}=0 & \mbox{ in } \Gamma_b \times [0,t_{\max}]
\end{array} \right. $$

To find the optimal values of $\{ \beta_i\}$ we used the gradient descent method.
This method is based on the observation that if the function $\displaystyle j(\mathbf {\beta} )$ is defined and differentiable in a neighborhood of a point $\displaystyle \mathbf {\beta^0}=\{ \beta_1^0, \cdots, \beta_N^0\}$, then $\displaystyle j(\mathbf {\beta} )$ decreases fastest if one goes from $\displaystyle \mathbf {\beta^0}$ in the direction of the negative gradient of $\displaystyle j$ at $\displaystyle \mathbf {\beta^0}$ , e.g. $-\nabla j(\mathbf {\beta^0} )=\left\{ \frac{\partial j(\mathbf {\beta^0} )}{\partial \beta_1}, \cdots, \frac{\partial j(\mathbf {\beta^0} )}{\partial \beta_N} \right\}$. It follows that for $\displaystyle \gamma $ small enough, the value of $j$ in $\displaystyle \mathbf {\beta^1} =\mathbf {\beta^0} -\gamma \nabla j(\mathbf {\beta^0} )$ is smaller than $\displaystyle j(\mathbf {\beta^0} )$.  We get a sequence $\displaystyle \mathbf {\beta}^{0},\mathbf {\beta}^{1},\mathbf {\beta}^{2},\dots$ such that  $\displaystyle \mathbf {\beta}^{n+1}=\mathbf {\beta}^{n}-\gamma _{n}\nabla 
j(\mathbf {\beta}^{n})$ and  $\displaystyle j(\mathbf {\beta}^{n+1})\leq j(\mathbf {\beta}^n)$, for all $n \in \mathbb{N}_0$. The convergence of this sequence to a local minimum of $j$ depends on the properties of $j$, (for example, $j$ convex and $\displaystyle \nabla j$ Lipschitz).

The value of $\gamma_n >0$ is different in every step. For each $n$, we searched for the greatest $\gamma_n \geq 0$ such that it minimizes $j(\mathbf {\beta}^{n}-\gamma \nabla j(\mathbf {\beta}^{n}))$. This is achieved combining a dichotomy method and a method that approximates $j$ by a parabola, for more information about this procedure, see page 51 of  \cite{Pi}. 
\section{Sensitivity analysis}\label{secc:sa}

In \cite{BD} Blackwell and Dowding analyzed the usage of sensitivity parameters in connection with the estimation of thermal properties in the heat conduction equation. They stated that although parametric investigations may be done, sensitivity parameters are rarely computed (see also \cite{BeLo2007}). Sensitivity parameters help to understand the parametric dependence of an
experiment and shape our experience and intuition for future cases .

In inverse problems, the sensitivity parameter is the partial derivative of the output function (in our case the temperature) with respect to a parameter being determined, which is $\partial T/ \partial p$ for a parameter $p$. Due to the general interest on the comparison of magnitudes for different parameters, a scaled (sometimes called ``modified'') sensitivity parameter is used:
\begin{equation} \label{eq:SAB}
 S_p := p \frac{\partial T}{\partial p}\,.
\end{equation}

Note that equation (\ref{eq:SAB}) has units of temperature for all parameters, therefore magnitudes for various parameters can be directly compared. 
Small sensitivity parameters, or general insensitivity, are beneficial when the parameters are not well quantified, such as materials with no characterized thermal properties. Then the parameter is not influential in the thermal response. To estimate a parameter, however, the measured response has to be sensible to that parameter. In this case, the scaled sensitivity parameters are desired to be larger in magnitude (compared to the representative temperature) and linearly independent (having different shapes). The more sensitive the temperature is, the more valuable the temperature measurements are. In a similar way, the estimation of multiple parameters requires that the sensitivity, or the effect on temperature of each parameter, is different or independent of one another for each parameter. If two parameters have similar effects on temperature, their individual influence is difficult to distinguish.

In this work we analyzed the sensitivity of the superficial temperature, for which we defined the modified  sensitivity parameter as follows:
\begin{equation} \label{eq:SA2}
 S_p (t) := \lim_{\Delta \to 0} p \; \frac{T_{p+\Delta}(X_{\max},t) -T_{p}(X_{\max},t)}{\Delta}\,,
\end{equation}

where $X_{\max}$ indicates the superficial boundary, and $T_{\alpha}$ is the solution of the differential problem (\ref{eq:ODEProb}), using the value $\alpha$ in the parameter $p$.

We used the method of finite differences \cite{BD} to determine the sensitivity parameters. First we solved the direct problem  (\ref{eq:ODEProb}) using the values of the parameters  $p=(p_1,p_2,...,p_i,..,p_n)$, and then we solved the direct problem again but with the following values: $p_\triangle=(p_1,p_2,...,p_i+\triangle p_i,..,p_n)$. Finally we approximated the sensitivity parameters as follows:

\begin{equation} \label{eq:SA3}
 S_{p_i} (t) \approx  p_i \frac{ T_{p_\triangle} (X_{\max},t) -T_p (X_{\max},t)}{\triangle p_i}\,.
\end{equation}

We remark that as every partial derivative, it is dependent not only on time, but also depends strongly in the values assumed by the parameters $p_i, \; i=1...n$.

This previous analysis works if the parameters are constant. In the case of $f$ which is variable in time, we introduced a new way of defining the sensitivity parameter of a variable parameter:

\begin{equation} \label{eq:sLs}
\left\{\begin{array}{c}
p=(p_1,p_2,...,f,..,p_n), \quad p_\triangle=(p_1,p_2,...,(1+\triangle).f,..,p_n)\\[2.5ex]
S_{f} ( t)\approx  \dfrac{ T_{p_\triangle} (X_{\max},t) -T_p (X_{\max},t)}{\triangle}
\end{array} \right. \\
\end{equation}

where $(1+\triangle).f$ represents the product of a real number and a function.

\begin{obs}
Note that in order to compute $S_{p_i}$ we need to work in a concrete problem, and therefore we will perform the sensitivity analysis for the biological application, where $f$ will be determined (Section \ref{secc:sensApl}).
\end{obs}


\section{Biological Application}\label{secc:ba}
It was previously demonstrated that the hamster cheek pouch is useful for the  study of tissue temperature affected by tissue superficial humidity \cite{SPIE2017, SantaCruz2011}. The hamster cheek pouch is widely used as a model of oral cancer and mucositis, an adverse side effect induced by several cancer therapies \cite{MontiHughes2015}. Our group is focused on the study of BNCT (Boron Neutron Capture Therapy), a binary treatment modality that can selectively target neoplastic tissue \cite{Coderre1999}. Particularly, we study BNCT therapeutic effect on tumors and BNCT induced mucositis in the hamster cheek pouch with a non-invasive complementary method called Dynamic Infrared Imaging (DIRI). This method is based on the observation of temperature changes under transient conditions associated with mass moisture transfer in the tissue-air interface of the pouch. 
In our previous studies, we described different temperature changes for normal and tumor tissue, and also for non-irradiated and irradiated pouches \cite{SPIE2017}. However, the study of the mass moisture transfer as a function of time was not quantified.   

\subsection{Dynamic Infrared imaging (DIRI) studies in the hamster cheek pouch}
Dynamic Infrared imaging (DIRI) is based on the acquisition of thermal images during transient processes, caused by sudden and sustained changes in surface temperature due to the application of a thermal stimulus (provocation test) that forces the neurovascular system to respond in order to maintain local and body temperature within normal parameters  \cite{Kellogg2006}. Other authors followed this concept, including our group \cite{SPIE2017,  SantaCruz2011, SantaCruz2009}, in different clinical research studies using thermography \cite{Arora2008, Hildebrandt2010, Pirtini2011, Bhavani2014}.  DIRI provides a non-invasively supplementary in vivo information potentially useful to characterize normal and pathological tissues and their response to cancer therapy. 

The biological model, experimental setup and procedures of the DIRI studies can be found in \cite{SPIE2017}. Briefly, a total of 61 hamsters were examined under DIRI protocol. Following an acclimatization period in the room, the animals were anesthetized and the pouch was everted using a plastic pipette held by hand. Thermal responses were measured using a FLIR T420 infrared camera, before, during and after the provocation test, namely, Transient Equilibrium Phase (TEP), Provocation Test (PT) and Recovery Phase (RP), respectively. The PT consisted of a mild air current applied at ambient temperature for about 120 seconds. The purpose of an air stimulus is to eliminate the initial moisture condition of the tissue so that, in the RP, we can focus on its thermal behavior and evaporation process that occur in response of the PT. In TEP and RP no air was applied, leaving the pouch exposed to ambient conditions without perturbations during approximately 280 and 400 seconds, respectively. 

\begin{figure}[!h]
\centering
\subfigure[]{ \includegraphics[height=5cm]{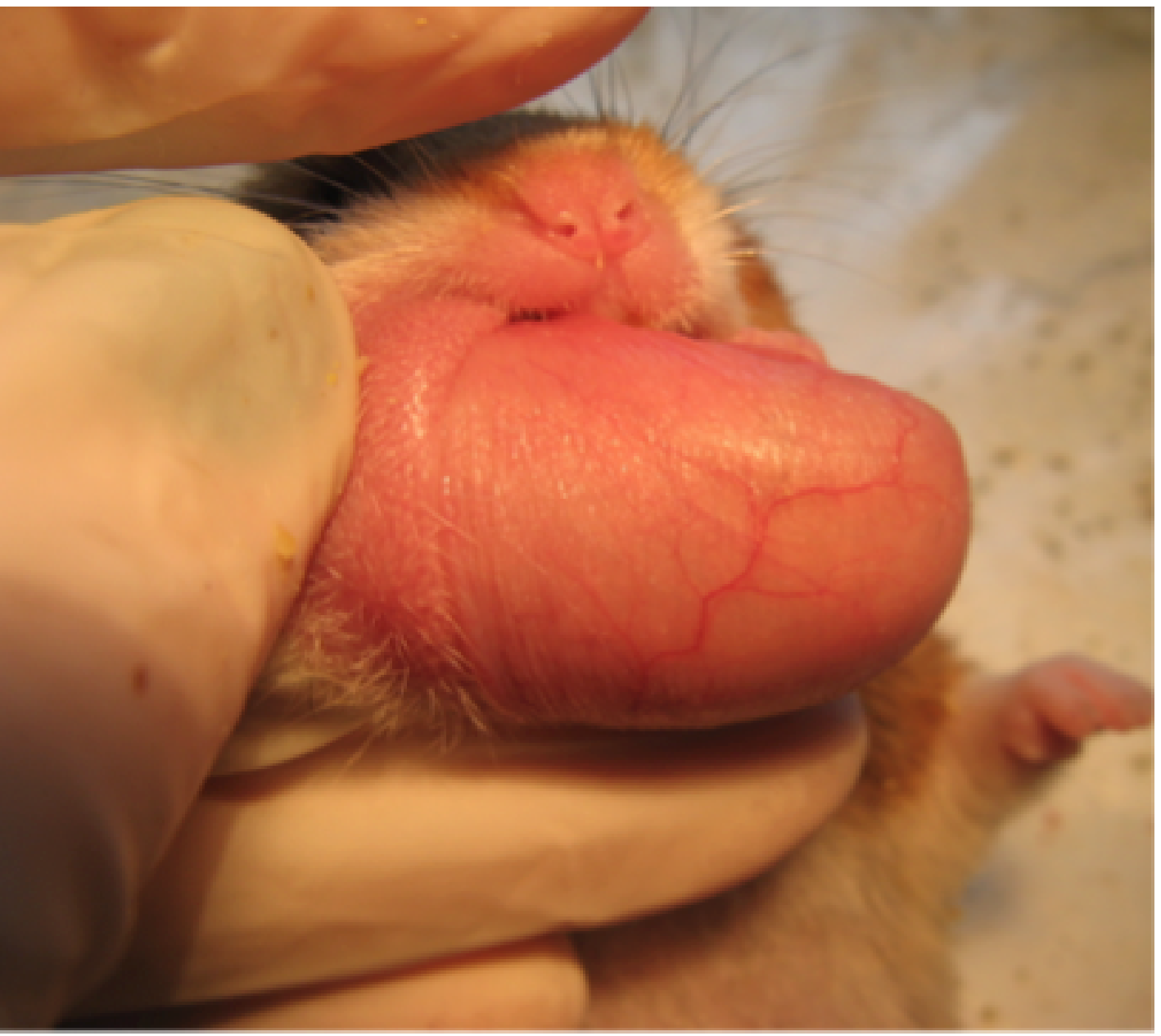}}
\subfigure[]{\includegraphics[height=5cm]{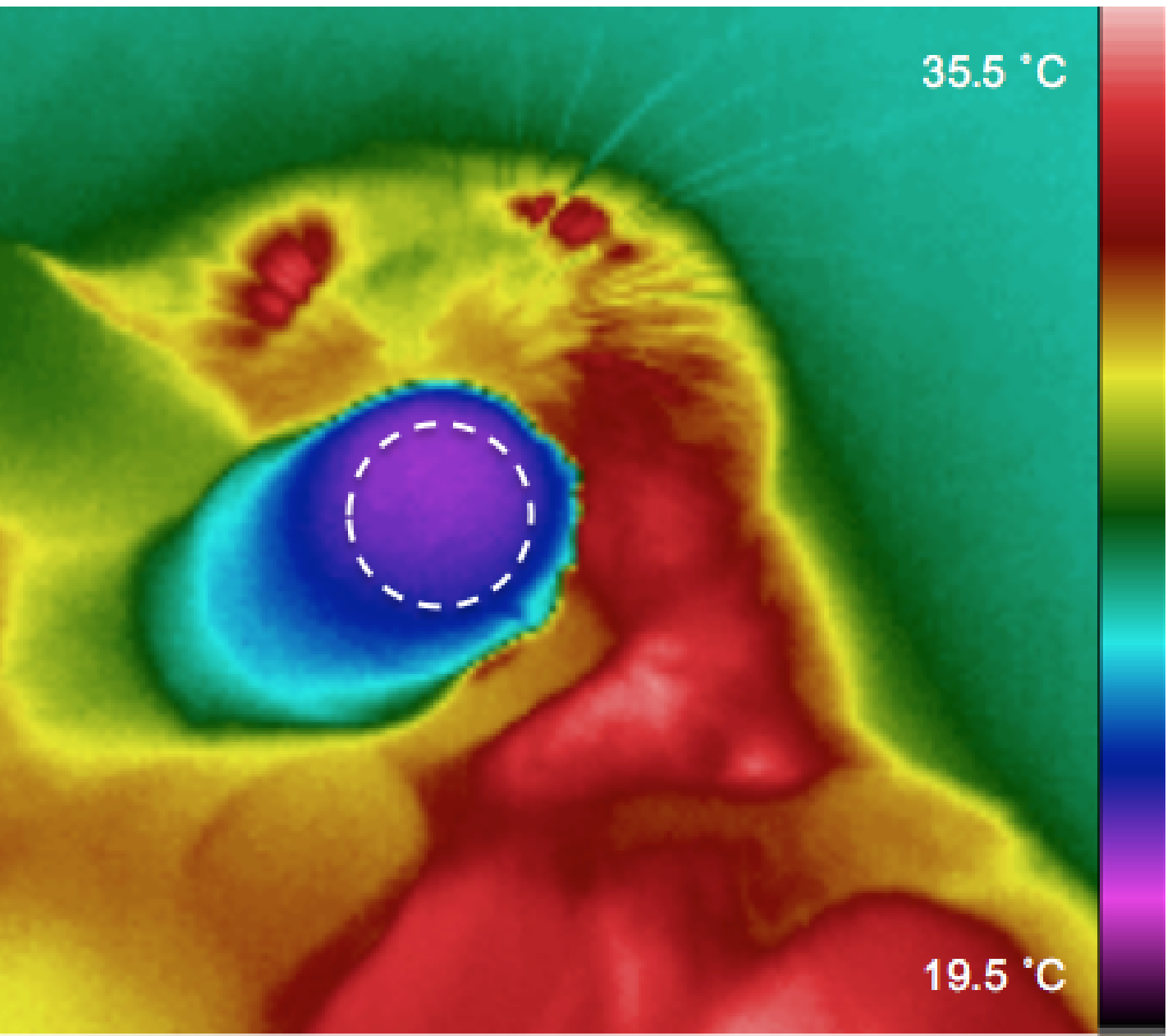}}

\caption{Example of a normal hamster cheek pouch (a) and a thermal image during the experiment. The region of interest (ROI), marked with a dashed line, is shown in the IR image (b).} \label{fig:roi}
\end{figure}

Figure \ref{fig:roi}(a) shows the normal hamster cheek pouch tissue. The measured temperature values during time were extracted from the thermal image (Figure \ref{fig:roi}(b)) and averaged in a user-defined region of interest (ROI) used to delineate the  normal tissue.

\subsection{Determination of the superficial temperature and superficial heat loss}
In this study, the differential equation model (\ref{eq:ODEProb}) was applied to model the hamster cheek pouch temperature, imposing a loss of heat in the superficial tissue, due to the water evaporation. For this simulation, we used the experimental temperature data as a function of time in a given ROI  \cite{SPIE2017}.

The heat transfer modeling in organs  has proposed numerous equations, studied by Pennes since 1948 \cite{Hu,Pe}. He suggested that the heat transfer rate between blood and tissue was proportional to the product of the volumetric perfusion rate and the difference between the arterial blood temperature and the local tissue temperature. Therefore the temperature of a tissue depends on the rate of blood perfusion, the metabolic activity and the heat conduction between the tissue and the environment. 
Taking into account the suggestion made by Pennes \cite{Pe}, Eq. (\ref{eqn:1}) of problem (\ref{eq:ODEProb}) takes the form:

\begin{equation} \label{eq:EDO}
\rho c \frac{\partial T (x,t)}{ \partial t }=k \Delta_x T +\omega_b \rho_b c_b (T_b-T)+q_m, \quad (x,t) \in \Omega \times [0,t_{\max}]
\end{equation}

\noindent where  $\rho$ ($\rho_b$) represents the tissue (blood) density, $c$ ($c_b$) is the tissue (blood) specific heat, $k$ is the thermal conductivity, $\omega_b$ is the blood perfusion coefficient, $q_m$ is the metabolic heat source and $T_b$ is the constant blood temperature. The boundary condition Eqs. (\ref{eqn:2}), (\ref{eqn:3}) and  (\ref{eqn:4}) are:


\begin{equation} \label{eq:BC}
 \left\{ \begin{array}{lll}
&-k \dfrac{\partial T}{\partial x}=h (T-T_{amb})+L(t), &   \mbox{in } \Gamma_u \times [0,t_{\max}]\\
 & T=T_D, &  \mbox{in } \Gamma_b \times [0,t_{\max}]\\
 & T(x,0) =  F_0(x) &\forall t \in [0,t_{\max}]\\
\end{array} \right. 
\end{equation}

Where,  $\Gamma_b=\{0\}$ represents the inner boundary and let $\Gamma_u=\{X_{\max}\}$  represents the superficial boundary. Here, $f(t) = L(t)$ ($W/m^2$) represents the superficial heat loss due to water evaporation, $h$ is the heat transfer coefficient  between the tissue and the air and $T_{amb}$ is the ambient temperature. We used the following initial temperature, that assures the continuity of the temperature values between the two boundaries  $\Gamma_b$ and $\Gamma_u$:

\begin{equation} \label{eq:F0}
F_0(x)=\dfrac{(T_0-T_{b})}{X_{\max}} x+T_{b}\,,
\end{equation}
 where $T_0$ is the initial superficial temperature measured. We used a linear function for simplicity.

Since thermal responses in the hamster cheek pouch were assessed before, during and after the application of a thermal stimulus (TEP, PT and  RP) \cite{SPIE2017}, these three different phases were modeled considering that the heat transfer coefficient $h$ assumed different values in each stage:

\begin{equation} \label{eq:h}
 h=h(t)= \begin{cases}
		     h_1 & \mbox{, if } \;0<t\leq 120\\
		     h_2 & \mbox{, if } \;120<t \leq 200\\
		     h_3 & \mbox{, if }\; 200<t \leq 400\\
                 \end{cases}
\end{equation}

The values used for $h_1,h_2$ and $h_3$ and others parameters are shown in Table \ref{tb:datosN2temp}.

\begin{table}[!h]
\centering
 \begin{tabular}{|ccccccc|} \hline
 $T_0$ & $T_{amb}$ & $T_b$ & $T_D$  & $h_1$ & $h_2$ & $h_3$ \\
 $^\circ C$ & $^\circ C$ & $^\circ C$ & $^\circ C$ &  $W/ m^2 \; ^\circ C$&  $W/ m^2 \; ^\circ C$&  $W/ m^2 \;^\circ C$\\ \hline
 28.715 & 28.5 &35 &$T_{amb}$ & 10 & 30 & 10 \\ \hline
 \end{tabular}
\caption{Thermal coefficients depending on the experiment.} \label{tb:datosN2temp}
\end{table}


To our knowledge, there are no published data related to the thermophysical properties (such as thermal conductivity, diusivity, etc.) of the hamster cheek pouch. In \cite{Poppendiek1966}, Poppendiek et al. suggested that tissues may be considered accurately for thermal analysis as being composed of water, protein and fat. Thus, for the thermal properties needed to compute Eq. (\ref{eq:SAB}) in our biological application, we only considered water and protein to calculate the thermal properties, shown in Table \ref{tb:datosN2}. The other parameters not related to the composition of the tissue were obtained from \cite{PirH}.


\begin{table}[!h]
\centering
 \begin{tabular}{|ccccccc|}
\hline 
 $k$ & $\omega_b$ & $\rho$ & $\rho_b$  & $c$ & $c_b$ & $q_m$ \\ 
 \scriptsize (W/m $^\circ$C) & \scriptsize (W/m$^3$ $^\circ$C) &  \scriptsize (Kg/m$^3$) & \scriptsize (Kg/m$^3$) & \scriptsize (J/Kg $^\circ$C) & \scriptsize (J/Kg $^\circ$C) & \scriptsize (W/m$^3$) \\ \hline
 0.445 &  0.0002 &  1200 & 1060 & 3300  & 3770 & 368.1 \\
\hline                 
\end{tabular}
\caption{Thermal coefficients used in the calculation.} \label{tb:datosN2}
\end{table}

\begin{figure}[!h]
\centering
\subfigure[Superficial temperature.]{\includegraphics[height=5cm,width=8cm]{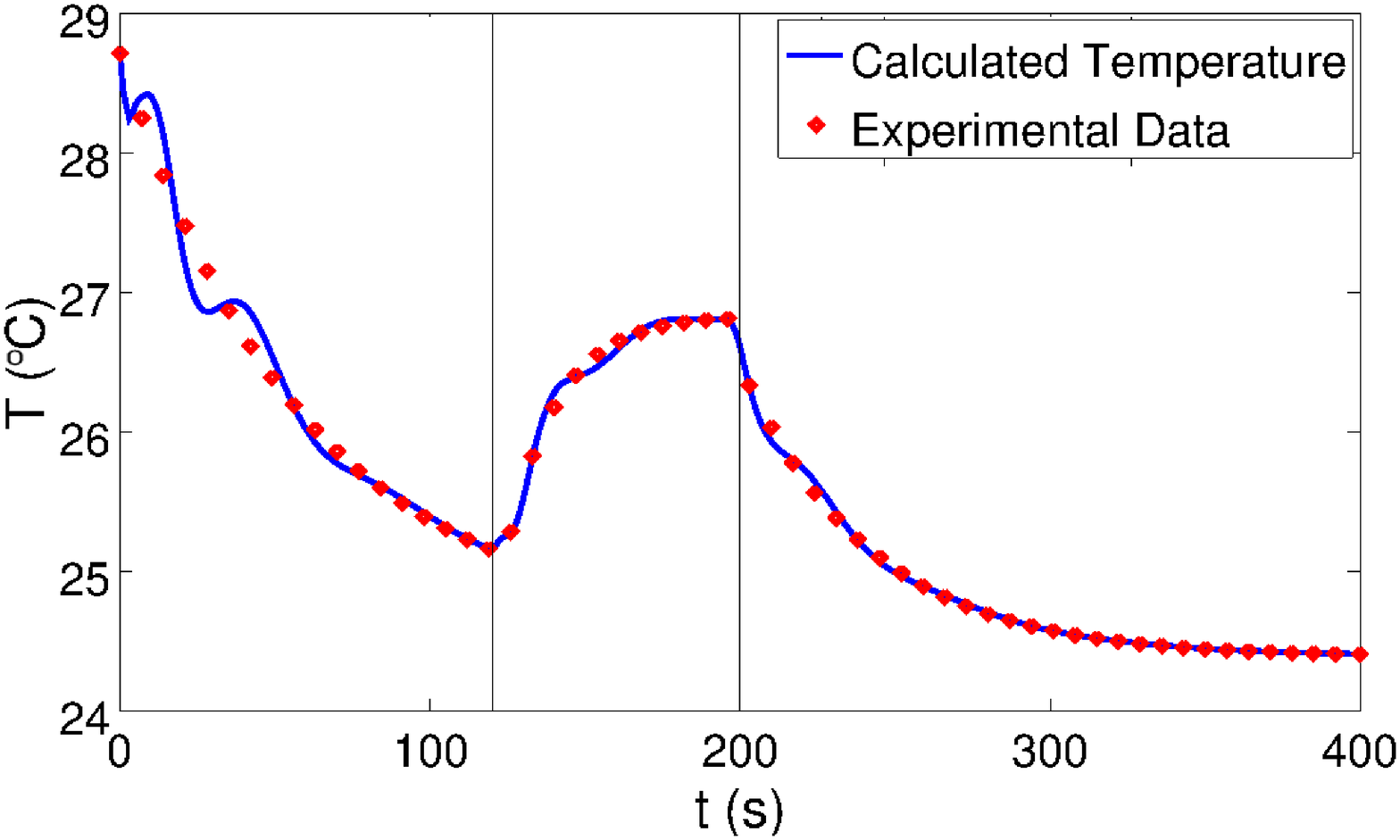}  }
\subfigure[Superficial heat loss.]{\includegraphics[height=5cm,width=8cm]{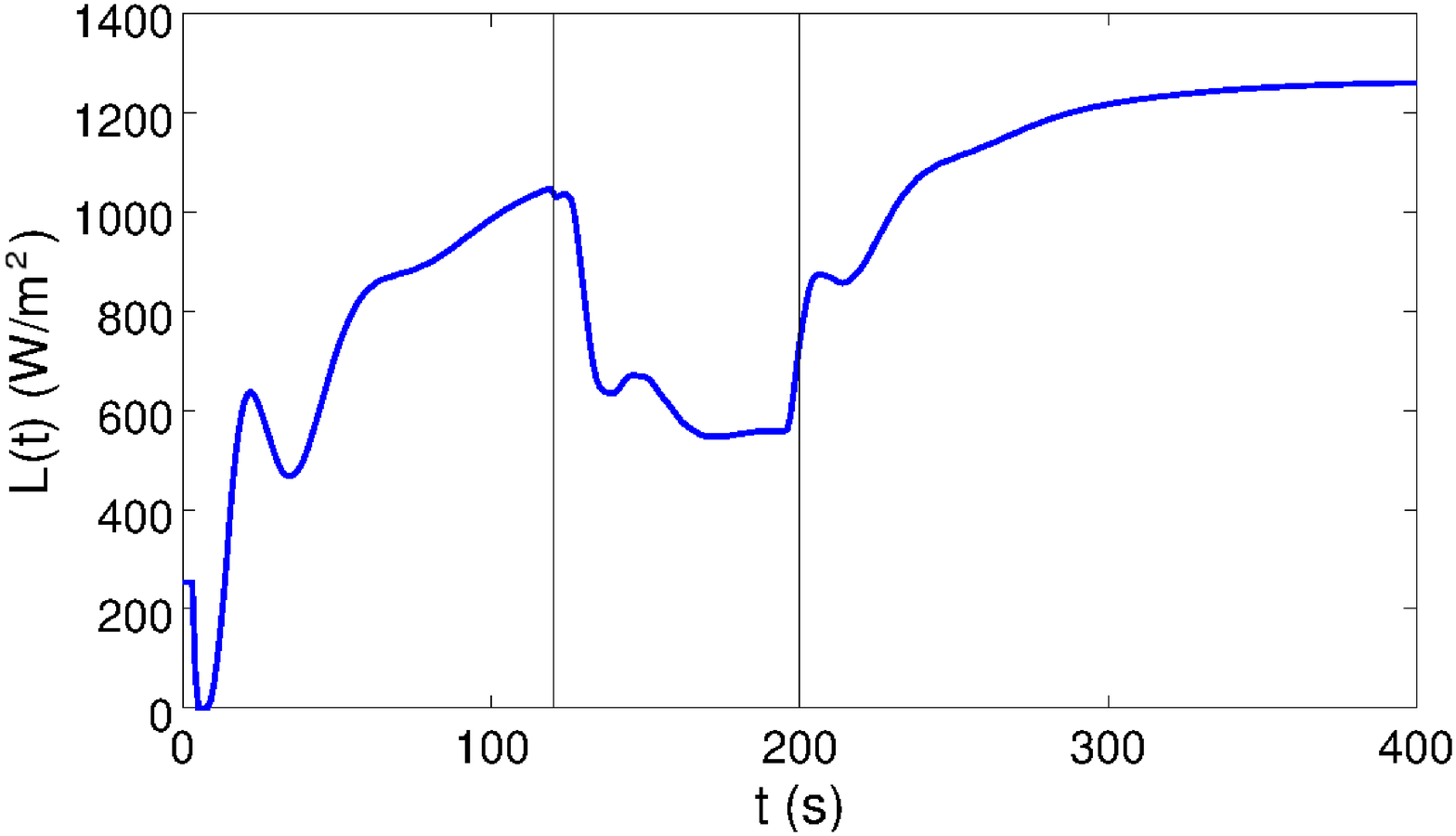}  }
\caption{(a) Normal hamster cheek pouch experimental and calculated thermal response as a function of time. (b) Superficial heat loss obtained through experimental data of figure (a). Vertical dashed lines separate the transient equilibrium phase (TEP), Provocation Test (PT) and Recovery Phase (RP).} \label{fig:sLs}
\end{figure}

Figure \ref{fig:sLs} shows the normal hamster cheek pouch experimental and calculated thermal response as a function of time. In Figure \ref{fig:sLs}(a) it can be seen the good approximation between the calculated  superficial temperature and the experimental data.  In particular, Figure \ref{fig:sLs}(b) depicts the convective coefficient obtained using  an initial constant coefficient $L(t) = 1000$ $W/m^2$, and 15 steps of the gradient descent method described in Section \ref{secc:detf}.

In Figure \ref{fig:sLs}(b),  we observed that at the beginning of each stage the superficial heat loss had oscillations. These oscillations can be seen also in the derivative of the functional (see Eq. (\ref{eq:E})), and therefore it seems that they are intrinsic of this differential problem. Besides, during the provocation test, the superficial heat loss decreases, due to the air stimulus that helps to eliminate the tissue superficial moisture.

\subsection{Sensitivity analysis} \label{secc:sensApl}
\begin{figure}[!ht]
\centering
\subfigure[Sensitivity of $T_{amb}$.]{\includegraphics[trim=20 0 20 10 0,clip=true,width=0.4\textwidth]{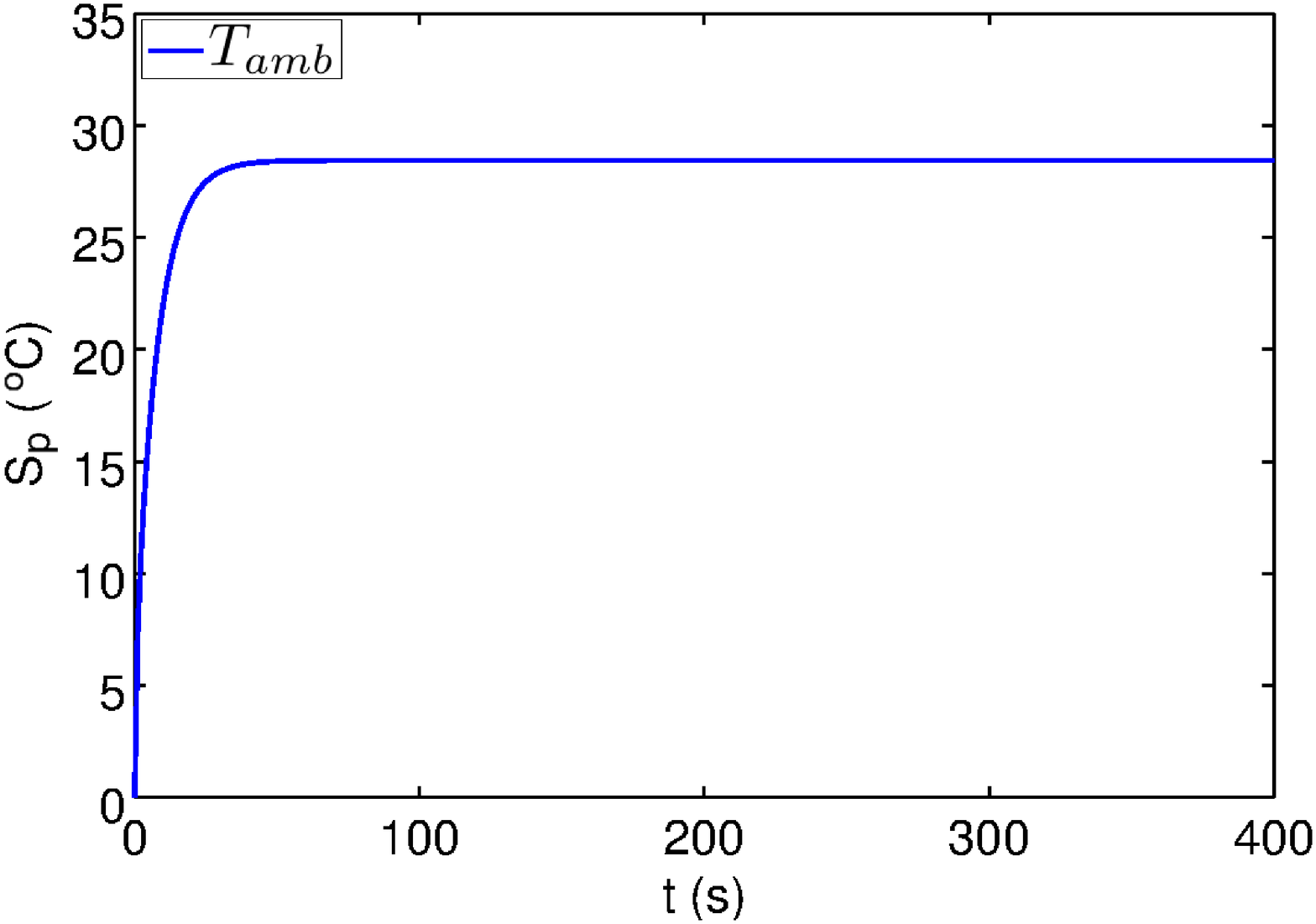}}
\subfigure[Sensitivity of $k$ and $L$.]{\includegraphics[trim=20 0 20 10 0,clip=true,width=0.4\textwidth]{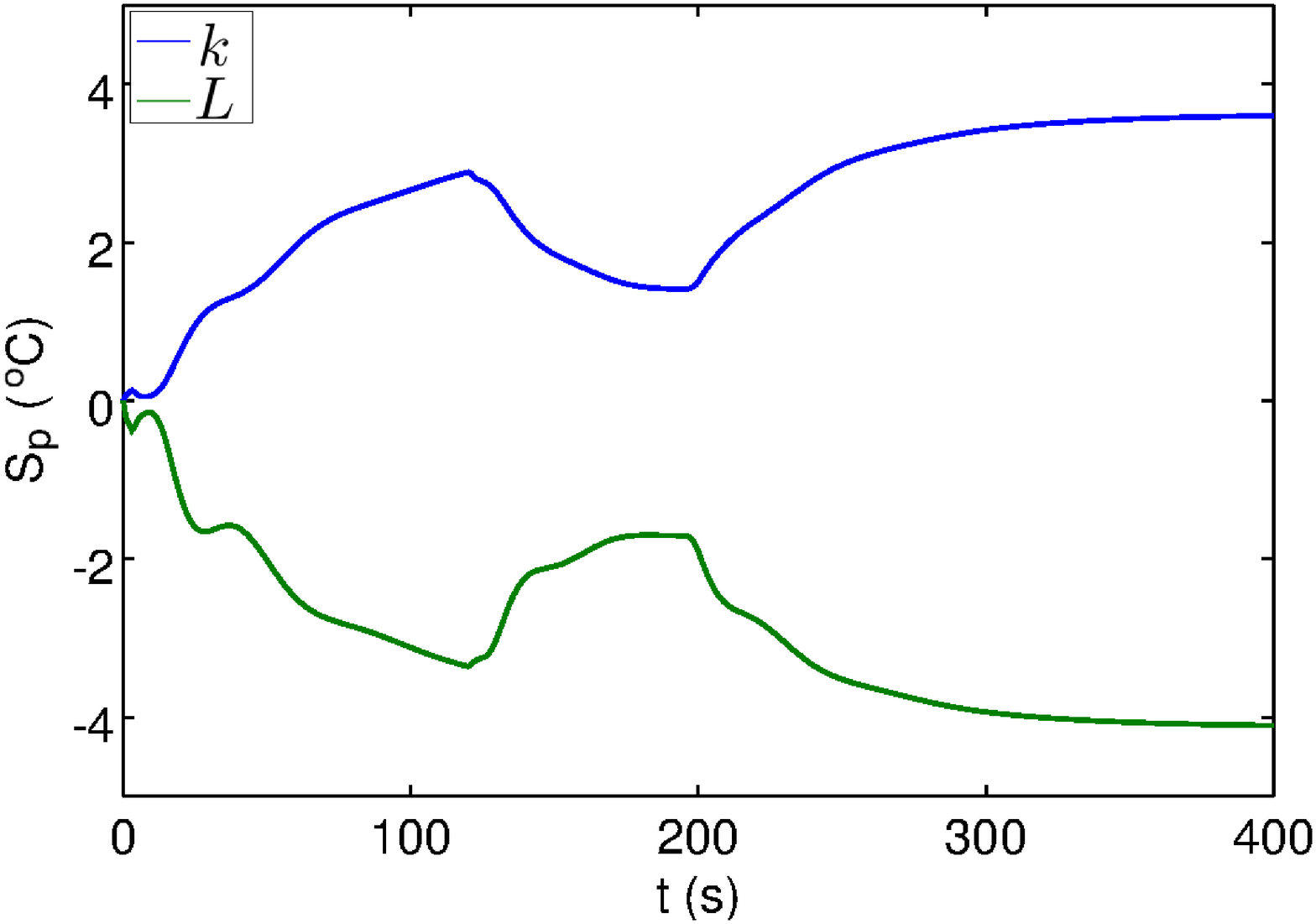}}
\subfigure[Sensitivity of $c$ (and $\rho$).]{\includegraphics[trim=20 0 20 10 0,clip=true,width=0.4\textwidth]{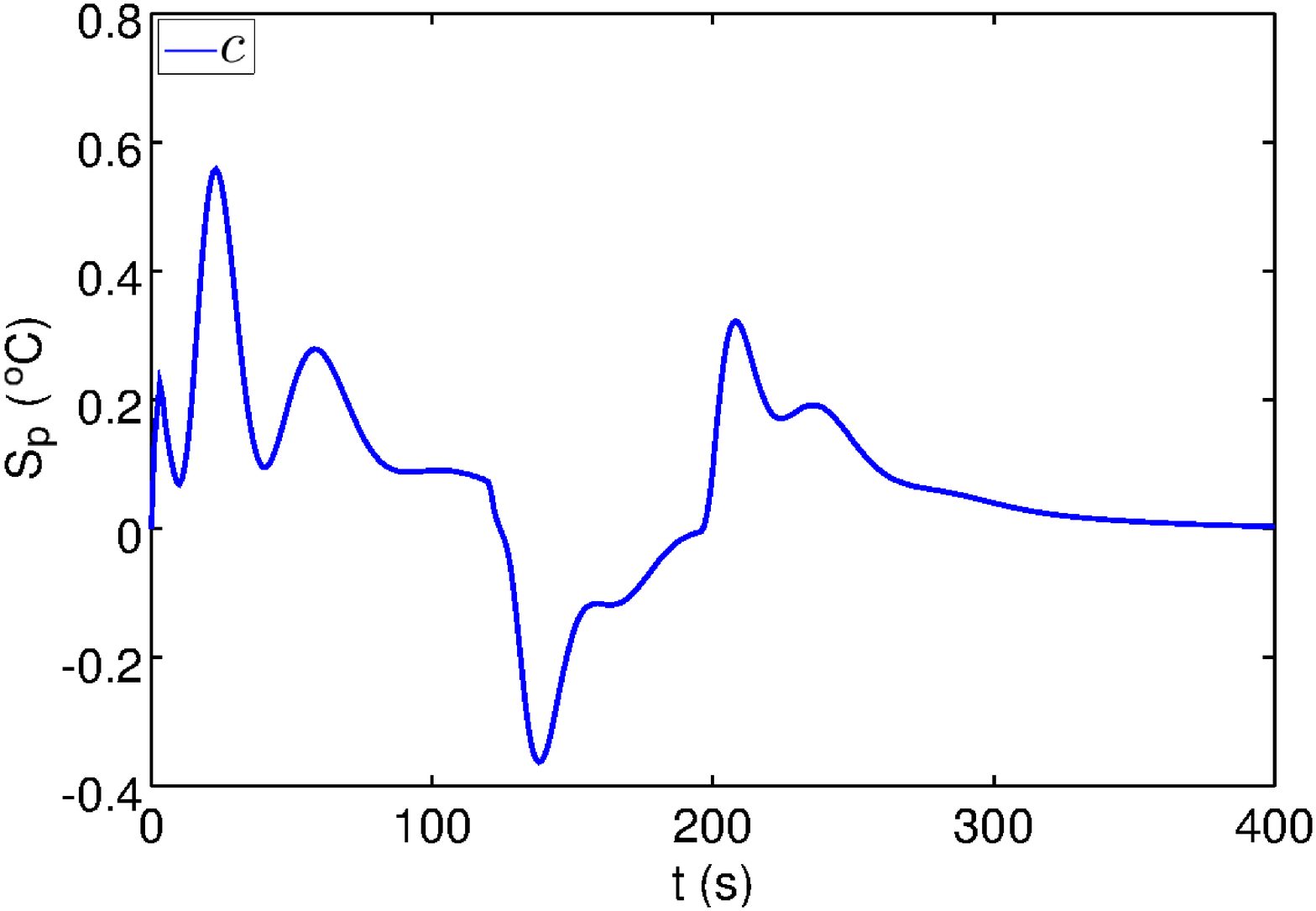}}
\subfigure[Sensitivity of $h$, $\omega_b$ and $T_b$.]{\includegraphics[trim=10 0 20 10 0,clip=true,width=0.4\textwidth]{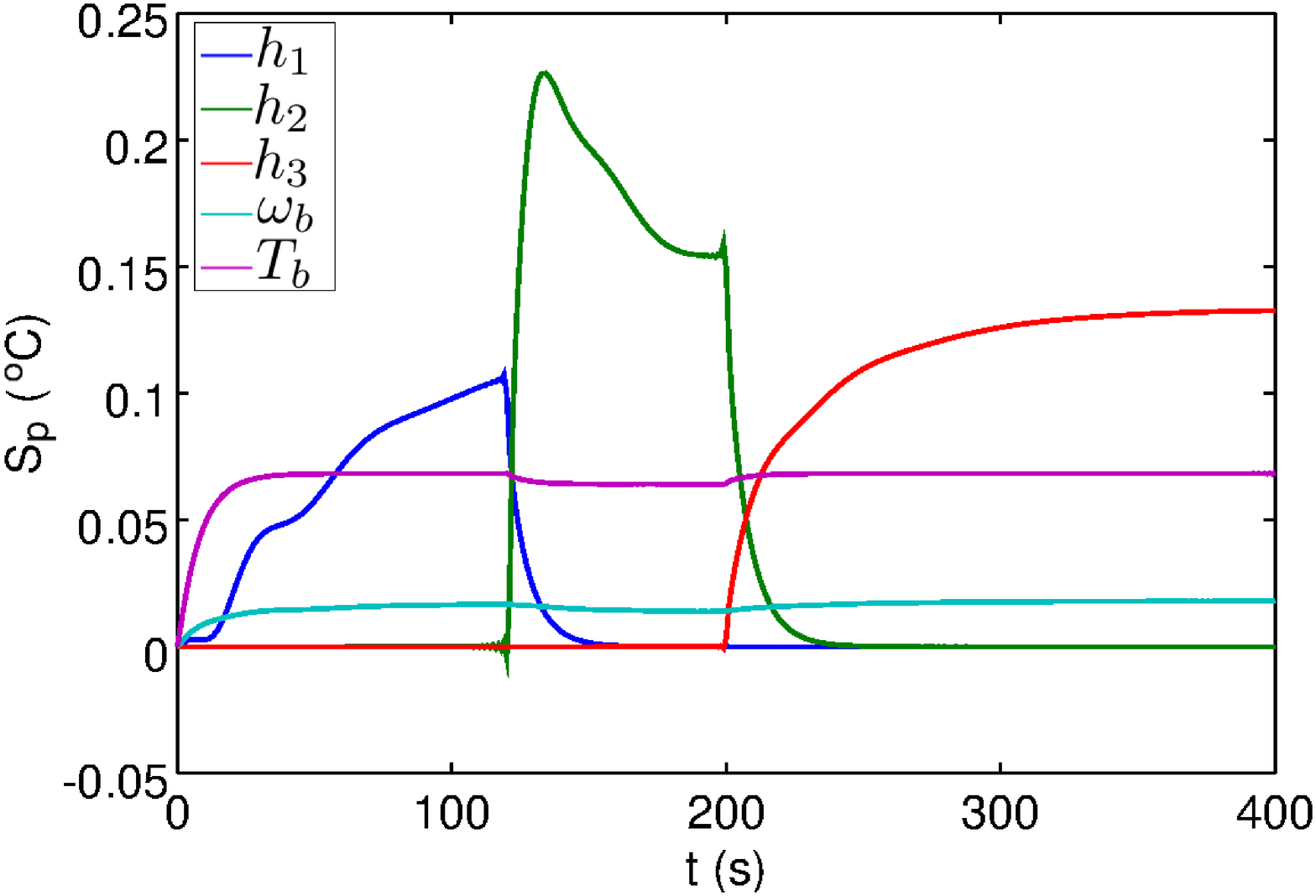}}
\caption{Scaled sensitivity as a function of time of different parameters.} \label{fig:sens} 
\end{figure}

Table \ref{tb:datosN2temp} and  \ref{tb:datosN2} show those parameters used for the sensitivity analysis. We also used the parameter $L(t)$ in Figure \ref{fig:sLs}(b), obtained in the previous section. The initial temperature is in Eq. (\ref{eq:F0}). Table \ref{tb:unidades} summarizes the physiological parameters considered, showing their units, and their position in the differential problem\footnote{The position in the differential problem, which was taken into account when the sensitivity was computed, changing the parameter $p$ to $p + \Delta$ whenever it appeared.}, necessary for the sensitivity analysis.  Note that some parameters, such as $\omega_b$ and $k$, appeared in more than one place in the differential problem. Figure \ref{fig:sens} shows different parameters sensitivities, grouped together depending on their order of the sensitivity. A summary of the order of the sensitivity parameters is shown in Table \ref{tb:sens}.

\small

\begin{table}[!htb]
\begin{center}
\begin{tabular}{|clp{5cm}l|} \hline
 & &  &\\
Symbol & Units & Represents & Position in the differential problem$^*$ \\\hline
 & &  &\\
$T(x,t)$ & $^\circ$C& temperature of tissue & DE, BC.\\
$\omega_b$&  1/$s$ & blood perfusion & DE: indep. term, heat source\\
$k$ & W/m $^\circ$C & tissue conductivity & DE: second order term\\
 & & & BC: external normal derivative\\
$\rho$ &  Kg/m$^3$ & tissue density & DE: temporal derivative\\
$\rho_b$ &  Kg/m$^3$ & blood density & DE: temporal derivative \\
$ c$ &  J/Kg $^\circ$C & specific heat of tissue & DE: temporal derivative\\
$ c_b$ &  J/Kg $^\circ$C & specific heat of blood & DE: temporal derivative\\
$\alpha$ & m$^2$/$s$ & diffusivity of tissue & DE: second order term\\
$\alpha_b$ & 1/$s$ & diffusivity of blood & DE: independent term\\
$q_m$ & W/m$^3$ & metabolic heat & DE: heat source\\
$Q$ & W/m$^3$ &  $\omega_b \rho_b c_b T_b+q_m$ & DE: heat source\\
$T_{amb}$ & $^\circ$C& ambient temperature & BC: convective condition\\
$T_b$ & $^\circ$C& arterial temperature & DE: heat source\\
&&& BC: Dirichlet condition\\
$T_D$ & $^\circ$C& inferior temperature & BC: Dirichlet condition\\
$T_0$ & $^\circ$C& superficial  temperature in $t=0$ & BC: Dirichlet condition\\
$F_0(t)$ & $^\circ$C& initial temperature & BC: Dirichlet condition\\
$h$ &  W/m$^2$ $^\circ$C & tissue-air convective coeff.& BC: convective condition\\
$L(t)$ &  W/m$^2$ & superficial heat loss &  BC: convective condition\\
$X$ & m & tissue depth & EDP: Domain\\
 \hline
\end{tabular}
\caption{ Physiological parameters: Units and symbols. $^*$DE: Differential equation, BC: Boundary conditions. } \label{tb:unidades}
\end{center}
\end{table}

\normalsize




\begin{table}[!ht]
\hspace*{5cm}\begin{tabular}{|cl|}
\hline Order of the sensitivity  & Parameters\\
10 & $T_{amb}$ \\
1 & $L, k$\\
$10^{-1}$ & $\rho, c$\\
$10^{-2}$ & $h, \omega_b, T_b$\\
$10^{-4}$ & $q_m$ \\
 \hline
\end{tabular}\\
\caption{Order of the sensitivity of the different parameters.} \label{tb:sens}
\end{table}

The first important observation is that the most sensible parameter is the ambient temperature ($T_{amb}$) (Figure \ref{fig:sens}(a)). This could be explained by the thinness of the tissue and the Dirichlet condition in the inner surface. Therefore, this parameter should be determined with the smallest error possible.

Secondly, we examined the linear dependence between parameters, studied in \cite{BA}, which could be established by analyzing the shapes of the sensitivity parameters. If two parameters are linearly dependent, this means that there are infinite possible solutions, which implies that there will be infinite local minimums and therefore the minimization problem will not converge.

Figure \ref{fig:sens}(b) shows that $L$ and $k$ are linearly dependent parameters, because their sensitivity are symmetric with respect to the horizontal axis. Therefore, although they have similar sensitivities, they should not be simultaneously determined. Figure \ref{fig:sens}(d) shows that $\omega_b$ and $T_b$ are linearly dependent parameters, because their sensitivities have the same behavior (increasing functions). Moreover, these parameters are linearly dependent with $h_1,h_2$ and $h_3$. Therefore to determine a parameter, of this order of sensitivity,  we should choose between $\omega_b$, $T_b$ and $\{h_1,h_2,h_3\}$.

Finally the least sensible parameter is the metabolic heat $q_m$, which may be justified by observing that the total heat source is $Q=\omega_b \rho_b c_b T_b+q_m$, and the predominant term is $\omega_b \rho_b c_b T_b$.


\pagebreak

\section{Conclusions}

We proposed a method for the determination of time dependent parameters using measured superficial temperatures in a conduction problem with evaporation, and a new way of computing a sensitivity analysis of a variable parameter. We applied this method successfully to a biological problem, modeling tissue temperature changes, under transient conditions, in the hamster cheek pouch. In this study we found a good approximation between the calculated superficial temperature and the experimental data. We performed a sensitivity analysis, which should be done whenever parameters are simultaneously determined. 
Based on temperature measurements and having calculated the loss of heat due to water evaporation at the superficial layer of the pouch, the sensitivity analysis determined which of the studied parameters were the most sensible to variations of the superficial experimental data. We found that ambient temperature should be measured with the smallest error, because it is the most sensible parameter in this problem. Moreover, we noted that the linear dependence between the conductivity and the superficial heat loss is not intuitive,  in contrast with the dependence between the blood perfusion and the blood temperature (an increase in either of them would result in a raise in the superficial temperature). Therefore, we conclude that a choice has to be made between determining the conductivity or the superficial heat loss. 

In previous studies, we observed that tumors and particularly a precancerous tissue bearing ulcers after BNCT had high superficial humidity \cite{SPIE2017}.  Thus, in future studies, the proposed mathematical model will be extended to explore the mass moisture transfer as a function of time in tumors and precancerous tissue in the hamster cheek pouch.

\subsection*{Acknowledgments}
This paper has been partially supported by the BNCT project of CNEA, and by CONICET.

\bibliographystyle{ieeetr}
\bibliography{biblio_2017}

\end{document}